\documentclass[12pt]{article}
\usepackage{amsfonts}
\usepackage{amsmath}
\usepackage{amsthm} 
\usepackage{dpscolor}
\usepackage[pdftex]{graphicx}

\newcommand{\NoBlackBoxes}{\global\overfullrule0pt}
\NoBlackBoxes

\numberwithin{equation}{section}

\newcommand{\myfootnote}{\footnote}

\begin{document}

\title{Hedging Against the Interest-rate Risk by Measuring the Yield-curve Movement}
\author{Zhongliang Tuo\myfootnote{The Pacific Securities Co.,Ltd. Email: tuozhl@itp.ac.cn.}
}

\date{Version: \today}

\maketitle

\abstract{
By adopting the polynomial interpolation method, we proposed an approach to hedge against the interest-rate risk of the default-free bonds by measuring the nonparallel
movement of the yield-curve, such as the translation, the rotation and the twist. The empirical analysis shows that our hedging strategies are comparable to traditional duration-convexity strategy, or even better when we have more suitable hedging instruments on hand. The article shows that this strategy is flexible and robust to cope with the interest-rate risk and can help fine-tune a position as time changes.
}

\section{Introduction}
\label{sec:Intro}

The determination of the interest-rate term structure is one important subject of the pricing models, the risk management,
the time value of money, hedge and arbitrage, \textit{et. al}. Many researches focus on the following five aspects: the formation of the term structure,
the statical models of the term structure, the micro analysis of the shape of the term structure, the dynamic models of the term structure and the empirical test of the dynamical models.

In capital market, hedgers, bond traders and portfolio managers concern more about the
anticipation of the changes in the term structure and the position of interest-rate based instruments. They try to estimate the movement of the interest rate and the risk exposure of the portfolio, and then they hedge against the risk by adjusting the position of instruments using some quantitative methods.

The first problem is how to estimate the movement of the interest rate. There are two
approaches to tackle this problem, one can be called the dynamics approach, and the other can be called the kinematics approach. The motivation of the first approach is that the
interest rate is determined by supply and demand of capital in the market, and one needs
to find out the impact factors (for example, some economic variables) that drive the movement of the interest rate, and one representative model is the multi-factor model with the econometric method and the principal component analysis~\cite{dewachter,ang,orphanides}. The motivation of the second approach is based on the observed properties of the interest rate, such as the mean reversion and the random fluctuation, and one use the equilibrium models~\cite{vasicek,cox1,cox2,rendkeman} or the no-arbitrage models~\cite{heath,hull,ho} to describe the movement of the yield-curve. The stochastic property of the interest rate may arise from the complicated impact factors yet unknown to us, so up to now, all existing models are only approximations, which would be invalid once the market environment changes.

The second problem is how to quantify the interest-rate risk once the yield-curve changes. A simple and widely used strategy is based on the concept of the duration~\cite{macauley,fisher}. Duration can be used to measure the sensitivity of the price to the change of the yield, and also can be used to calculate the hedge ratios. Redington~\cite{redington} proposed a method to immunize the bond portfolio against the parallel movement of the term structure by using the duration. But this method gives a sensible risk measure only if the yield-curve shifts in the parallel manner, thus the duration approach should be improved if the change of the yield-curve is nonparallel.

Nonparallel movement is more realistic in the real market. Many observational data indicates that there are two types of nonparallel movement, slope change and curvature change. For example, the term structure may become steep or flat, and the changes of the two sides may be different from the change of the middle, which is called butterfly shift. Many researchers have payed attention to the nonparallel movement before. Garbade~\cite{garbade} discussed the immunization method if the slope of term structure changes. Litterman and Scheinkman proposed a three-factor approaches by quantifying the level, the slope and the curvature of the term structure~\cite{litterman}, which has been widely used and generated by many researchers. Chambers and Carlet~\cite{chambers} introduced the concept of multiple duration, which they called duration vectors. This method is developed by Ho~\cite{ho1}, who introduced the concept of key-rate durations based on the interest rate on the maturity date. Even though these methods are helpful for estimating the interest rate risk, they are less helpful for predetermining the trade that should be made to hedge against the risk. Because of simplicity and tractability, the duration immunization method is still favored by many market participants and other traders.

By adopting the polynomial interpolation method, we propose a method that can measure the interest rate risk and hedge against the risk. This method preserves the concept of duration and takes into consideration of various movements of the yield curve, such as the translation, the rotation and the twist. One can also generalize to other cases in which more complicated evolution behaviors happens to the yield-curve, if one has suitable number of hedging instruments on hand.

This paper is organized as follows. In the next section, we introduce some main characters of the interest-rate term structure and its movement properties. In Sec. 3, we introduce our method to describe the changes of the yield curve, and then propose a dynamical method to immunize a single bond or a portfolio. In Sec. 4, we show the empirical test of our strategy and make comparisons with other methods proposed. The conclusion is present in Sec. 5.

\section{Statistical properties of interest rate term structure and immunization}
\label{sec:IRTS}

Many motivations of modeling the term structure dynamics arise from the empirical observations of the interest-rate term structure. Some important movement properties of the
interest-rate term structure are summarized below~\cite{bouchaud,rebonato,cont}:

1. Mean reversion: This behavior has resulted in models where interest rates are modeled as stationary processes.

2. Smoothness in maturity: This property should be viewed more as a requirement of market operators, which means that the yield curves do not present highly irregular profiles with respect to maturity. This is reflected
in the practice of obtaining implied yield curves by smoothing data points using
splines.

3. Irregularity in time: The time evolution of individual forward rates (with a fixed time to maturity) are very irregular.

4. Principal components: Principal component analysis of the term structure deformation indicates that at least two factors of uncertainty are needed to model term
structure deformation. In particular, forward rates of different maturities are
imperfectly correlated.  The shapes of these principal components are stable across time periods and markets.

5. Humped term structure of volatility: Forward rates of different maturities are
not equally variable. This hump is always observed to be skewed towards smaller maturities.
Moreover, though the observation of a single hump is quite common~\cite{Moraleda}, multiple
humps are never observed in the volatility term structure.

We show the movements of the yield curve both with time (t) and maturity (T) in Figure 1. The data are selected by the Wind Financial Terminal~\myfootnote{http://www.wind.com.cn.} from the China Securities Index, which contains 3-year daily spot rate of the treasury bond, including the maturity of 6 month, 1 year, 2 year, 3 year, 4 year, 5 year, 6 year, 7 year, 8 year, 10 year, 15 year, 20 year. Figure 1 demonstrates the consistency of the term structure with the properties summarized above. The upper graph shows the evolution behaviors of the spot
rate of different maturities. The below graph shows the yield curves at different time. Table 1 demonstrates that the spot rates of different maturities are correlated at different level, and the correlation coefficients are all larger than 0.57.

\begin{figure}[!h]
\begin{center}
\includegraphics[width=0.7\textwidth,height = !]{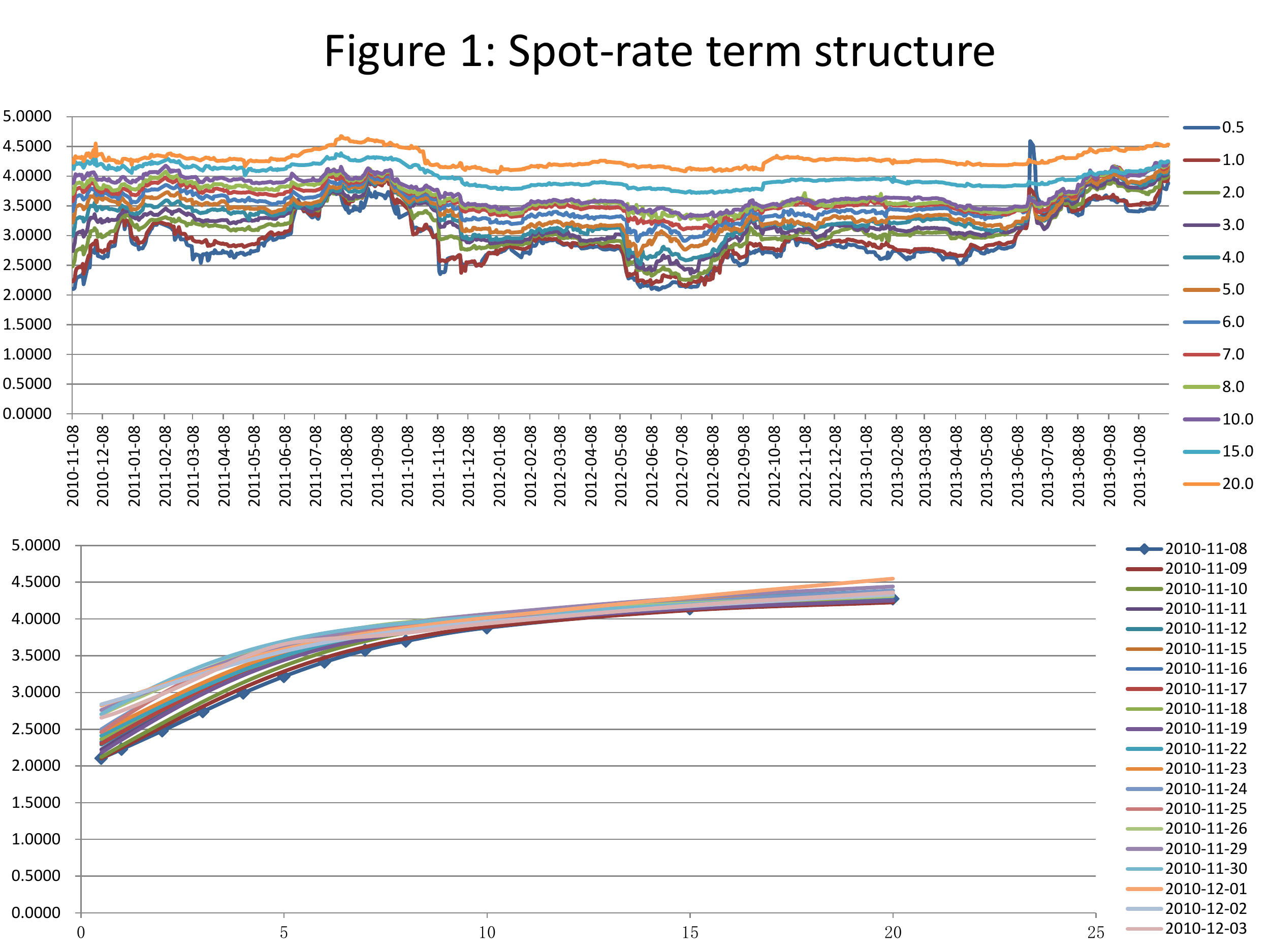}
\label{fig:irts}
\end{center}
\end{figure}

\begin{figure}[!h]
\begin{center}
\includegraphics[width=0.7\textwidth,height = !]{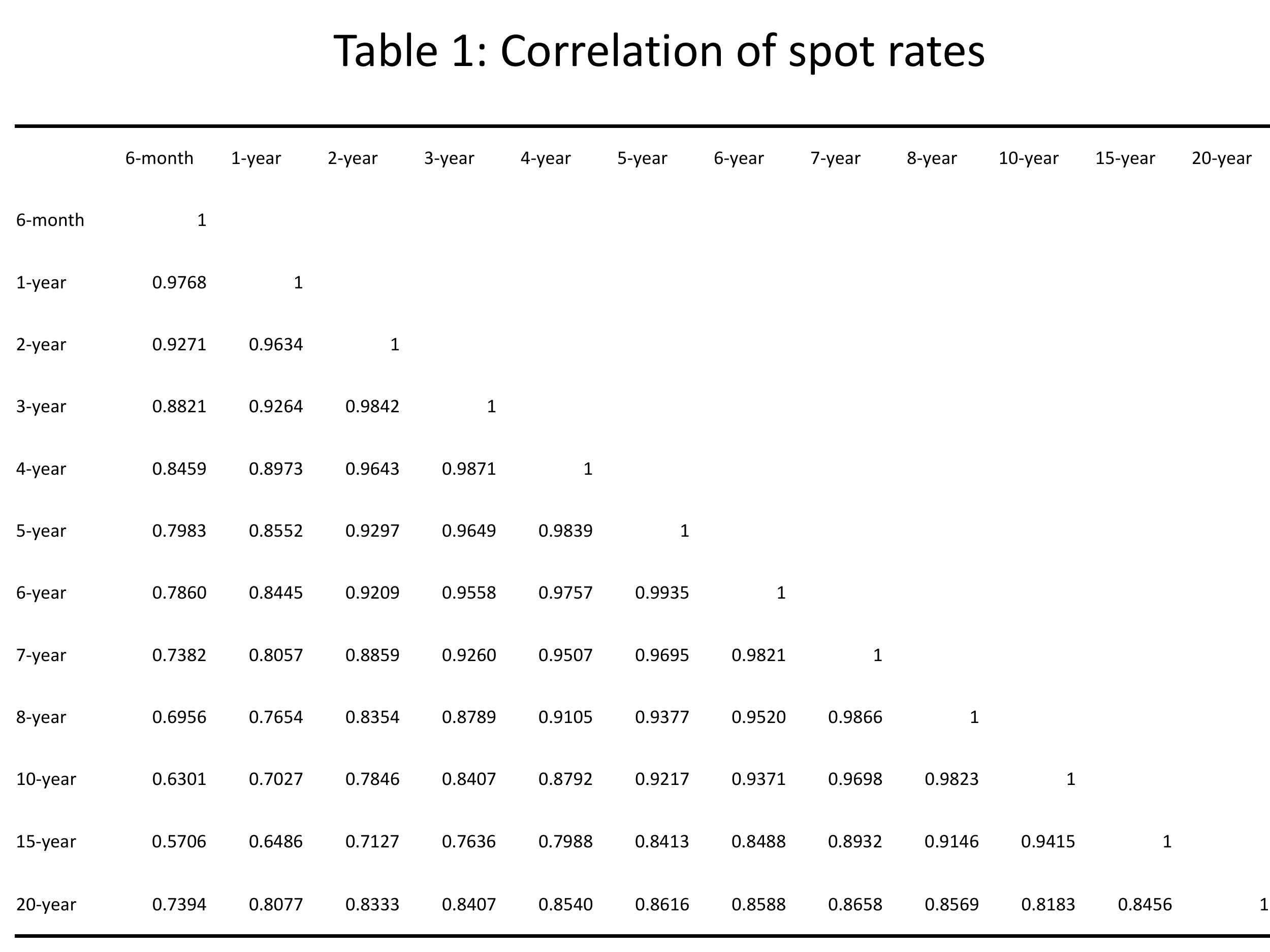}
\label{fig:irts}
\end{center}
\end{figure}

For the 6-month spot rate, the regression analysis demonstrates that it is not a random-walk or stationary process, and that it shows obvious serial correlation and unit-root characteristics, which is different from the former summary. This property may be generated in an inefficient market~\cite{Pesando}, and may cause estimation bias~\cite{ball}. Instead of discussing the reason for this, we focus on the hedging strategy against the interest rate risk.

Consider the following case: One wants to hedge a single bond with another one instrument against the interest rate. Suppose that he holds one bond $B$ with price $P$, amount $N$, duration $D$, and in order to hedge against the interest-rate risk, he sells an appropriate amount of standard hedging instrument $B_A$, for example a future contract or a benchmark bond, with price $P_A$, amount $N_A$ and duration $D_A$. The total value of the combination is $V=NP+N_AP_A$, which should be independent of the yield movement $\Delta y$. Then one ontains

\begin{equation}
N_A=-\frac{N\Delta P}{\Delta P_A}.
\label{duration}
\end{equation}

Specifically, if the movement $\Delta y$ is parallel or infinitesimal, the hedge ratio can be calculated as $N_A=-\frac{NPD}{P_AD_A}$, which is the ordinary duration-based hedging ratio. But for more general and realistic cases, one has to completely evaluate $\Delta P$ as
\begin{equation}
\Delta P=P(Y+\Delta Y)-P(Y).
\end{equation}

There are two problems lie before us, the first is how to express $\Delta Y$, and the other is how to completely evaluate $\Delta P$ once the yield curve changes.
Some researches focused on the first problem by calculating $\Delta Y$ with various approximations (for more details, see references~\cite{heiko,agca,crack}), other researches introduced duration-based
approaches to solve the second problem, such as the traditional duration-convexity, the exponential duration and the discrete duration (see references~\cite{livingston,bajo}).

From mathematical point of view, the most precise solution is to accurately express $\Delta Y$, and then completely calculate $P(Y+\Delta Y)$. This method means that one needs large number of hedging instruments to cover the interest-rate risk. It is time consuming and unrealistic, and it will generate new risks such as liquidity risk and basis risk. And on the other hand, some researchers find that higher-order principal components show increasingly oscillating profiles in maturity and the variances associated to these principal components decay quickly \cite{bouchaud,rebonato}. As a result, using large number of hedging instruments to cover the interest-rate risk may not be so efficient.

Unlike the traditional duration approach, we propose an method which allows for non-parallel movement of the yield-curve. This method does not rely on historical data, and one can easily adjust the hedging position depending on the market situation, which is flexible and not time consuming. The main object of this stratedgy is to hedge against the interest-rate risk with less instruments but with higher accuracy. We limit the number of hedging instruments
to 3 or less.

\section{The model}
\label{sec:model}

The interest-rate term structure is actually a curve in 3-dimensional space, which has two freedom degrees represented by two free parameters $(t,\, T)$, where $t$ is the time (such as date) and $T$ is the maturity, and $Y(t,\,T)$ is not static but evolutive with time. This curve shows smoothness in maturity but irregularity in time, so it is difficult to express it as an analytical formulation.

Because of its smoothness, we choose a segment of the yield-curve between $T_A$ and $T_B$ ($T_A<T_B$), which are the maturities of two hedging instruments, respectively. $T_B-T_A$ can not be too large, because under common conditions, a portfolio manager would unlikely use a long-maturity bond to hedge a short-maturity bond, which will increase the liquidity risk and basis risk. In China's treasury-bond future market, the maturity of the deliverable bond is between 4-7 year. So, we can safely use interpolation method to express the yield curve between $[T_A,\,T_B]$. In the following, we apply the physical concept $translation,\,rotation,\,twist$ to describe the movement of the yield-curve.

\begin{figure}[!h]
\begin{center}
\includegraphics[width=0.55\textwidth,height = !]{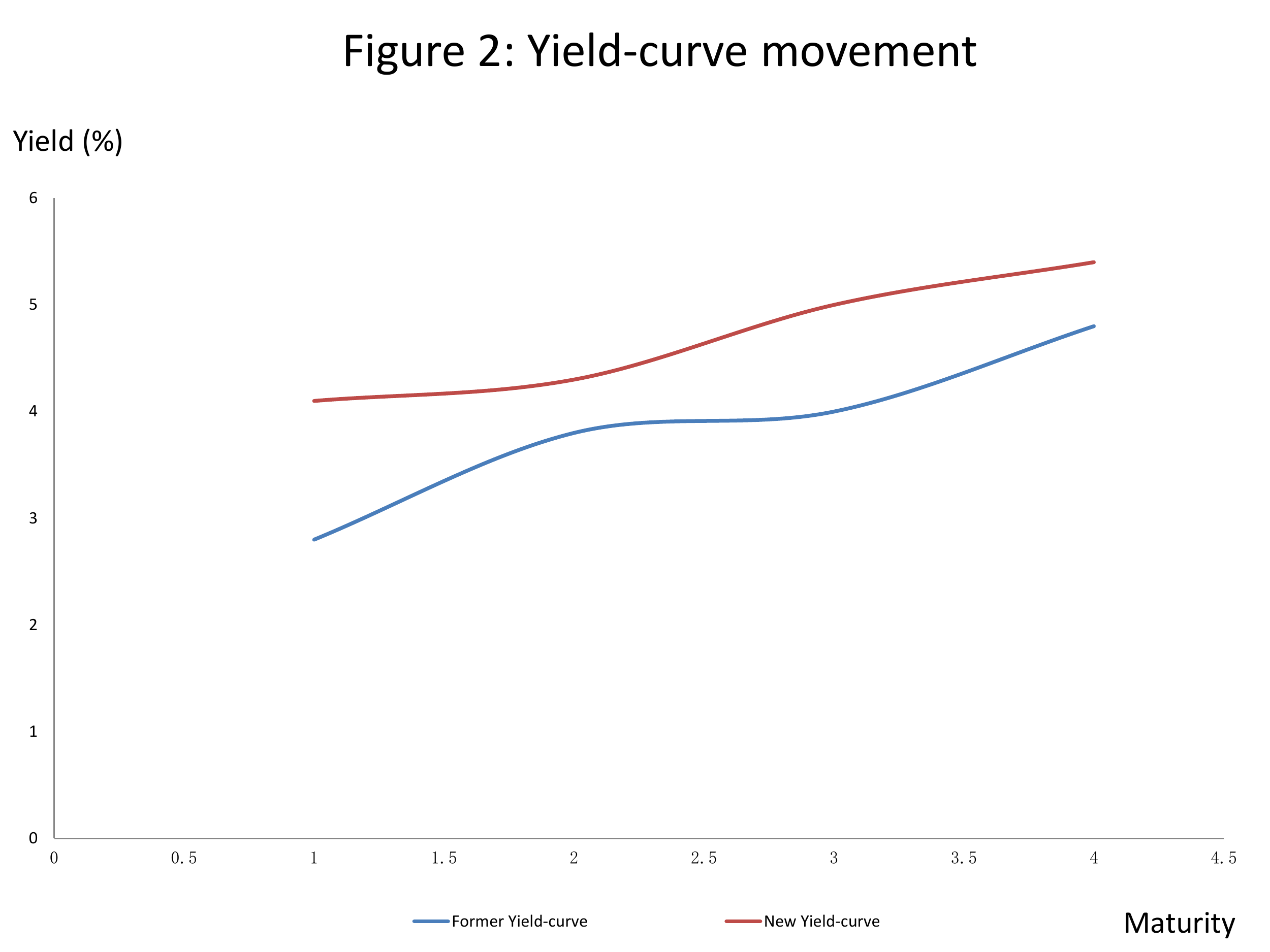}
\label{fig:movement}
\end{center}
\end{figure}

Suppose the yield-curve between $[T_A,\,T_B]$ can be approximated as a polynomial $Y(t)$, and we need at least cubic polynomial in order to quantify the $twist$.

\begin{equation}
Y(T)=\alpha+\beta T+\gamma T^2+\lambda T^3.
\label{y}
\end{equation}

where the coefficients $\alpha,\,\beta,\,\gamma$ are constants determined by the hedging instruments we choose. The $translation,\,rotation,\,twist$ can be expressed by $\alpha$, the first-order derivative and the second-order derivative (or the curvature $K(T)=\frac{F''(T)}{(1+F'^2(T))^{3/2}}$).

We can see from Figure 2 that the $translation,\,rotation,\,twist$ can represent the common movement of the yield-curve very well, but they are not co-moving with each other. As a 2-dimension curve, it is the other variable $t$ that determines the change of each kind of movement, in another word, $translation,\,rotation,\,twist$ should be related to different functions of $t$, respectively. So, we can express $\Delta Y(t,\,T)$ as follows,

\begin{equation}
\Delta Y(t,\,T)=a(t)+b(t)F'(t)+c(t)F''(T).
\label{deltay}
\end{equation}


where, $a(t),\,b(t),\,c(t)$ are independent time-dependent functions. Modifying $a(t),\,b(t),\,c(t)$ is equivalent to changing the level, the slope and the curvature of the yield-curve, respectively. Equation~\ref{y} means that we need 3 standard instruments to hedge against the movement of the yield-curve. When using 2 hedging instruments, we need to drop the third term of Equation~\ref{y}.

Once we obtains the expression of $\Delta y$, the other problem left is how to evaluate $\Delta P$. No matter which approach is adopted (such as the exponential duration, discrete duration)~\cite{livingston,bajo}, one needs two more hedging instruments, thus leads to more complex hedging strategy. Since there is no robust evidence that the exponential duration approach or the discrete duration approach overmatches the duration-convexity approach, we will adopt the traditional duration-convexity approach to calculate $\Delta P$.


\begin{equation}
\Delta P=P(-D\Delta Y+\frac{1}{2}C\Delta Y^2).
\label{deltap}
\end{equation}

Suppose we hold $N$ bond with price $P$, maturity $T$, duration $D$, convexity $C$, and if we has one suitable hedging instruments on hand, with price $P_A$, maturity $T_A$, duration $D_A$, convexity $C_A$, then the most convenient and effective strategy is the duration strategy, with the hedge ratio $N_A=-\frac{NPD}{P_AD_A}$. This means that we can only cover the parallel movement risk of the yield-curve with only one hedging instrument.

If we hold two suitable hedging instruments, the situation begins to change. Since we have three instruments, we can determine three parameters in Equation~\ref{y}, so we need to ignore the third term. Accordingly, we have two kinds of hedging strategies as follows,

\begin{itemize}
  \item Ignoring the third term of Equation~\ref{y}, we obtain the following equation:

\begin{eqnarray}
NPD(a(t)+b(t)(\beta+2\gamma T)+c(t)\gamma)+N_AP_AD_A(a(t)+b(t)(\beta+2\gamma T_A)+c(t)\gamma)  \nonumber \\
+N_BP_BD_B(a(t)+b(t)(\beta+2\gamma T_B)+c(t)\gamma) = 0.
\end{eqnarray}

Thus, the following equations should be fulfilled in order to sufficiently hedge against the movement of the yield-curve,

\begin{eqnarray*}
NPD+N_AP_AD_A+N_BP_BD_B& = & 0,\\
NPDT+N_AP_AD_AT_A+N_BP_BD_BT_B & = &0.\end{eqnarray*}

Solving these equations, we arrive at the hedge ratios $N_A,\,N_B$ as follows,

\begin{eqnarray*}
N_A& = & -\frac{NPD}{P_AD_A}\frac{T_B-T}{T_B-T_A},\\
N_B& = & -\frac{NPD}{P_BD_B}\frac{T-T_A}{T_B-T_A}.
\label{compare1}
\end{eqnarray*}

We find that this result is the same as the result in~\cite{heiko}. But our results are more general and flexible if we have suitable number of hedging instruments. In the following, we call this approach as the quadratic approach. This approach means that we only consider the translation and rotation of the yield-curve, so we need two hedging instruments.

  \item Taking use of Equation~\ref{deltap}, which means that we adopt the duration-convexity approach (equally, $\Delta Y=a(t)$), we obtain the following equations:

\begin{eqnarray}
NP(-Da(t)+1/2Ca^2(t))+N_AP_A(-D_Aa(t)+1/2C_Aa^2(t))  \nonumber \\
+ N_BP_B(-D_Ba(t)+1/2C_Ba^2(t))  = 0.
\end{eqnarray}

Thus, the following equations should be fulfilled,

\begin{eqnarray*}
NPD+N_AP_AD_A+N_BP_BD_B& = & 0,\\
NPC+N_AP_AC_A+N_BP_BC_B & = &0.\end{eqnarray*}

Accordingly, the hedge ratios can be calculated as follows,

\begin{eqnarray*}
N_A& = & \frac{NP(C_BD-CD_B)}{P_A(C_AD_B-C_BD_A)},\\
N_B& = & \frac{NP(-C_AD+D_AC)}{P_B(C_AD_B-C_BD_A)}.
\label{compare2}\end{eqnarray*}

\end{itemize}

If we hold three suitable hedging instruments, we can fully determine Equation~\ref{y}. Taking use of Equation~\ref{y}, and following the same procedure, we obtain the hedge ratios for the three hedging instruments:
\begin{eqnarray*}
N_A& = & -\frac{NPD}{P_AD_A}\frac{(T-T_C)(T-T_B)}{(T_B-T_A)(T_C-T_A)},\\
N_B& = & -\frac{NPD}{P_BD_B}\frac{(T-T_C)(T-T_A)}{(T_B-T_A)(T_B-T_C)},\\
N_C& = & -\frac{NPD}{P_CD_C}\frac{(T_B-T)(T-T_A)}{(T_C-T_A)(T_B-T_C)}.
\label{compare3}
\end{eqnarray*}

where, we have set $T_A<T_C<T_B$. It is clear that this condition will not impact the result. In the following, we call this approach as the cubic approach.

Once the method to hedge a single bond is known, we can easily calculate the hedge ratios for a portfolio with $n$ bonds. Suppose that the maturity of each bond $T_i$ lies between $T_A$ and $T_B$. The amount, price, maturity, duration and convexity of theportfolio are $N,\,P,\,T,\,D,\,C$, respectively, which can be expressed as follows,

\begin{eqnarray*}
NP& = & \sum_{i=1}^{n}n_iP_i,\\
T&=& \max (T_i),\\
D& = & \frac{\sum_{i=1}^{n} n_iD_i}{\sum_{i=1}^{n}n_i},\\
N_C& = & \frac{\sum_{i=1}^{n} n_iC_i}{\sum_{i=1}^{n}n_i}.\\
\label{portfolio}\end{eqnarray*}

Combining the quadratic approach and the cubic approach with the above equations, we can use our model to hedge the bond portfolio.

In the next section, we will analyze the ability of our model in hedging against the interest-rate risk. We will make comparison of the duration approach, the quadratic approach, the duration-convexity approach and the cubic approach, respectively. To illustrate the results, we carry out an empirical study. The representative bond and the standard hedging instruments are actively traded treasury bonds selected from the Wind Financial Terminal.

\section{Comparison analysis}
\label{analysis}
In order to compare the hedging effect of different methods proposed in Section~\ref{sec:model}, we select 4 representative treasury bonds and use the daily data from 2007-06-04 to 2008-06-04. In practice, one tends to choose hedging instruments whose maturities are close to that of the representative bond or portfolio. For example, one would prefer to choose instruments with zero-year to 2-year maturity to hedge the portfolio with short maturity.

The representative treasury bonds and the maturity, price, modified duration, and convexity of each bonds on the starting date (June 4th, 2007) are listed in Table 2, which are actively transacted in Shanghai Stock Exchange.

\begin{figure}[!h]
\begin{center}
\includegraphics[width=0.6\textwidth,height = !]{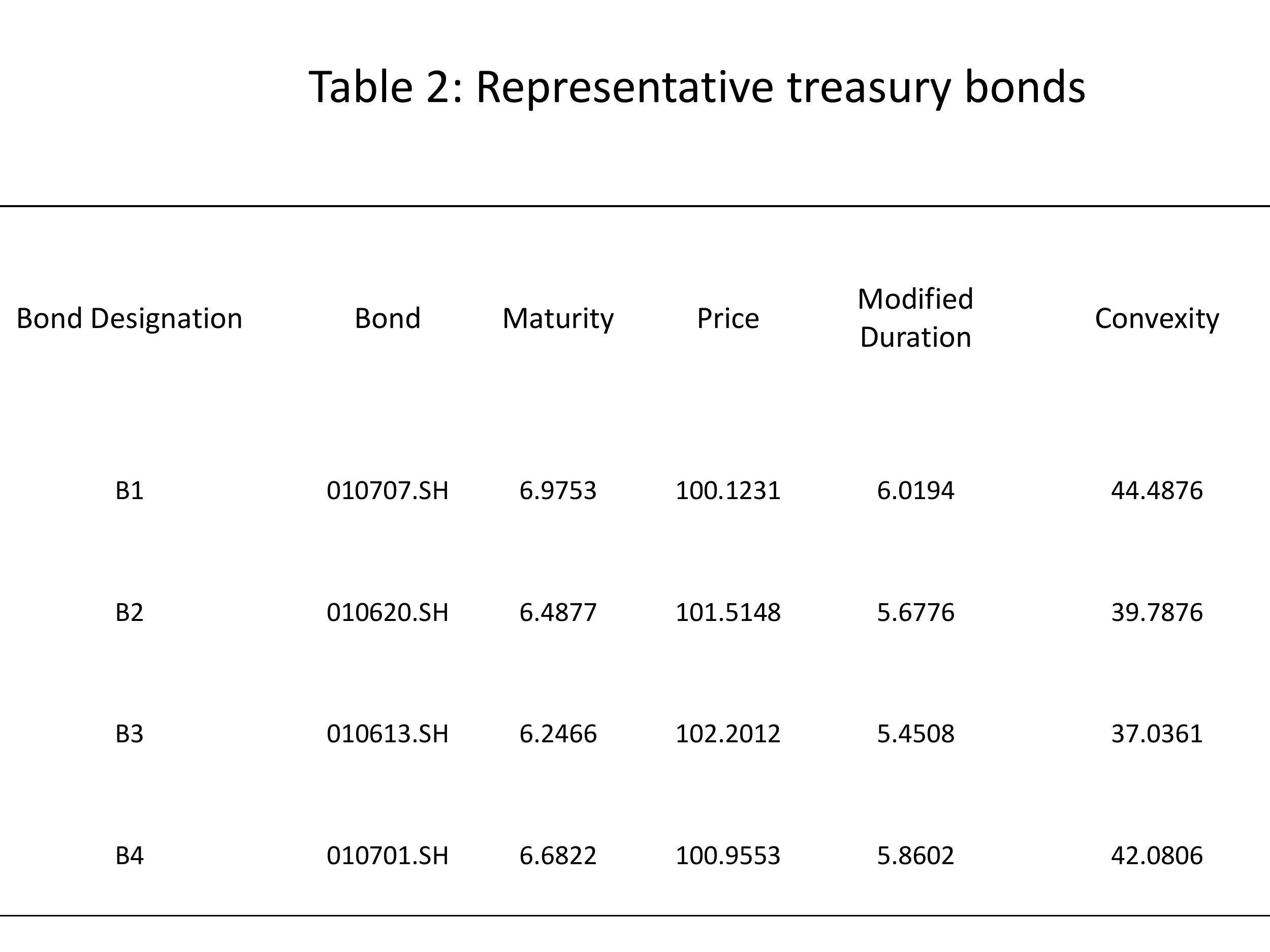}
\end{center}
\end{figure}

In the following, we will compare these hedging strategies by monitoring the daily profit-loss under each strategy. We suppose that the bond we hold is B2 with $N=100$, and the hedging instruments are B3 and B1, with the amount $N_A$ and $N_B$, respectively. When using three hedging instruments, we also add in B4, with the amount $N_C$. As a comparison, we also use the duration approach to hedge against the interest-rate risk. The results of these strategies are shown in Figure 3. It is obvious that the hedging strategies proposed in this paper are much more effective than the duration approach. But we also find that if the maturity drops to below about 6-month, these approaches can not hedge against the interest-rate risk so effectively, which means that these strategies lose efficacy when hedging against the ultra-short-term interest-rate risk. This may be caused by the high volatility and irregularity of the ultra-short-term interest rate.

\begin{figure}[!h]
\begin{center}
\includegraphics[width=0.8\textwidth,height = !]{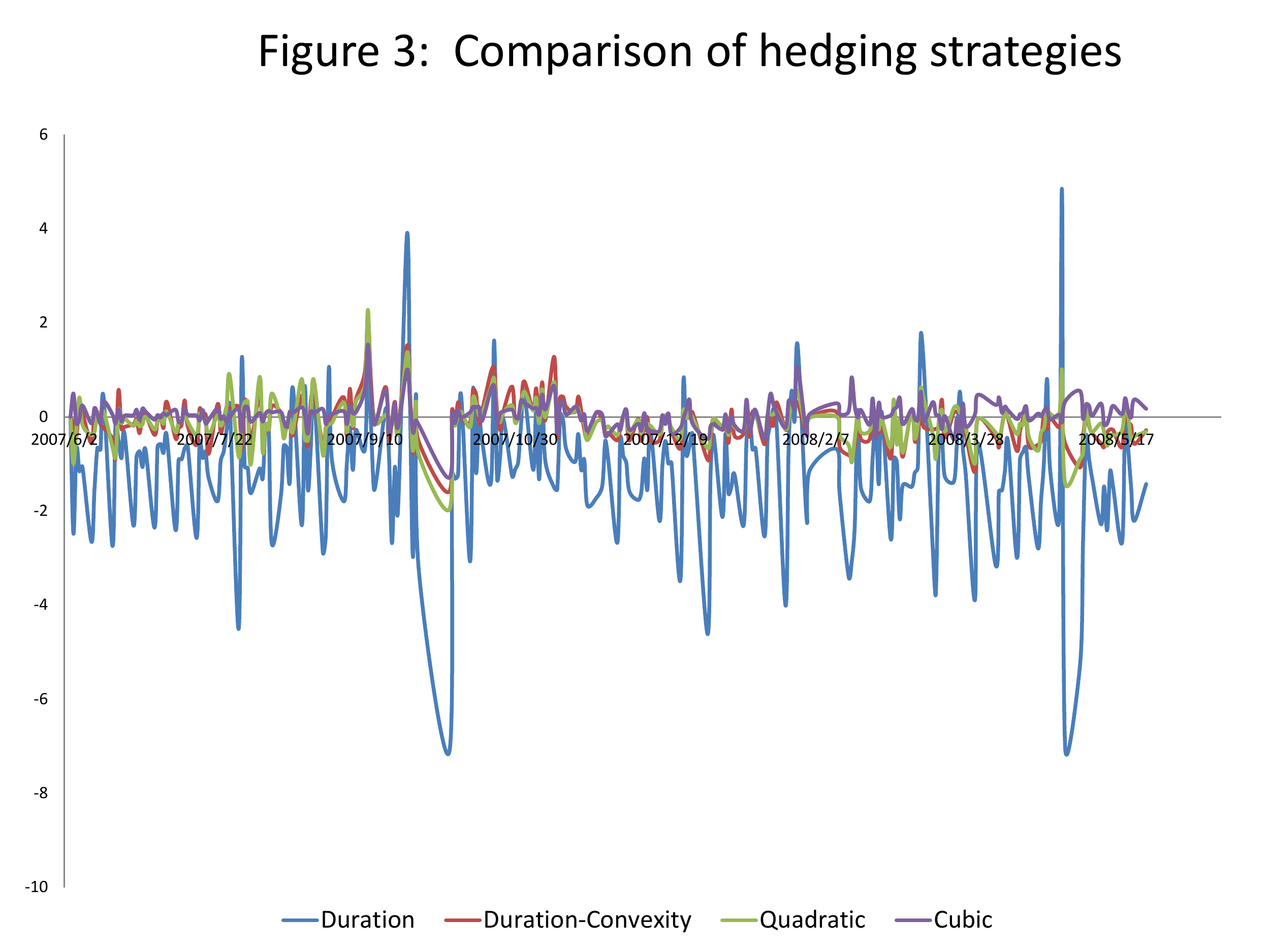}
\label{hedge1}
\end{center}
\end{figure}

We also compare the quadratic approach and the traditional duration-convexity approach, both of which contain two hedging instruments. The result is shown in Figure 4. We can see that both strategies are comparable when hedging against the interest-rate risk. But in some period (for example, the period around December, 2007), the quadratic approach performs better than the duration-convexity approach.

\begin{figure}[!h]
\begin{center}
\includegraphics[width=0.8\textwidth,height = !]{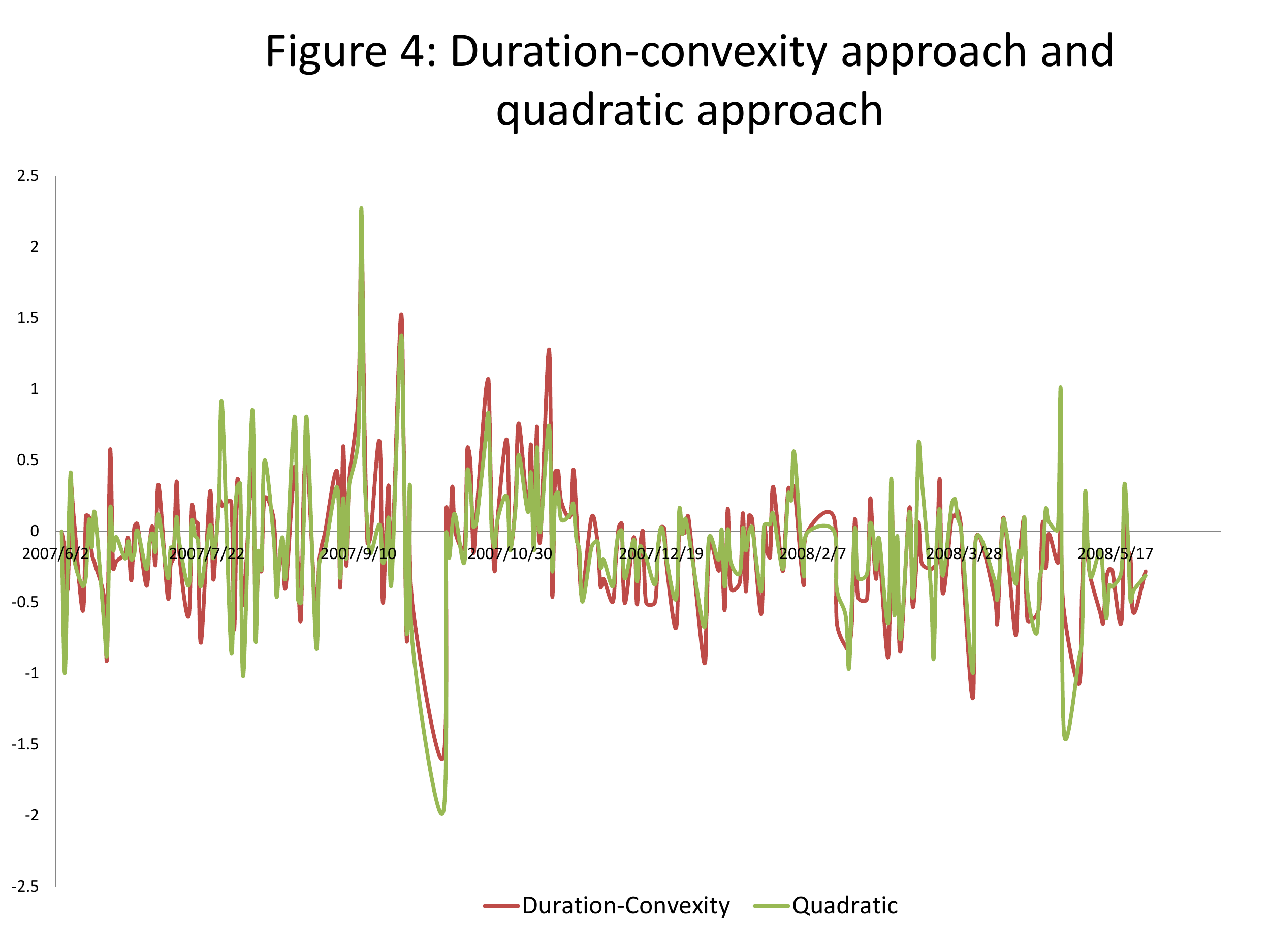}
\label{hedge2}
\end{center}
\end{figure}

Next, we will compare the quadratic approach and the cubic approach. The difference between these strategies is that the latter takes into consideration of the twist of the yield-curve. Figure 5 shows that the cubic approach performs obviously better than the quadratic approach. One reason for this is that we consider more information about the movement of the yield-curve, and the other reason is that we add B4 to immunize B2 against the interest rate, whose maturities are closer to each other.

\begin{figure}[!h]
\begin{center}
\includegraphics[width=0.8\textwidth,height = !]{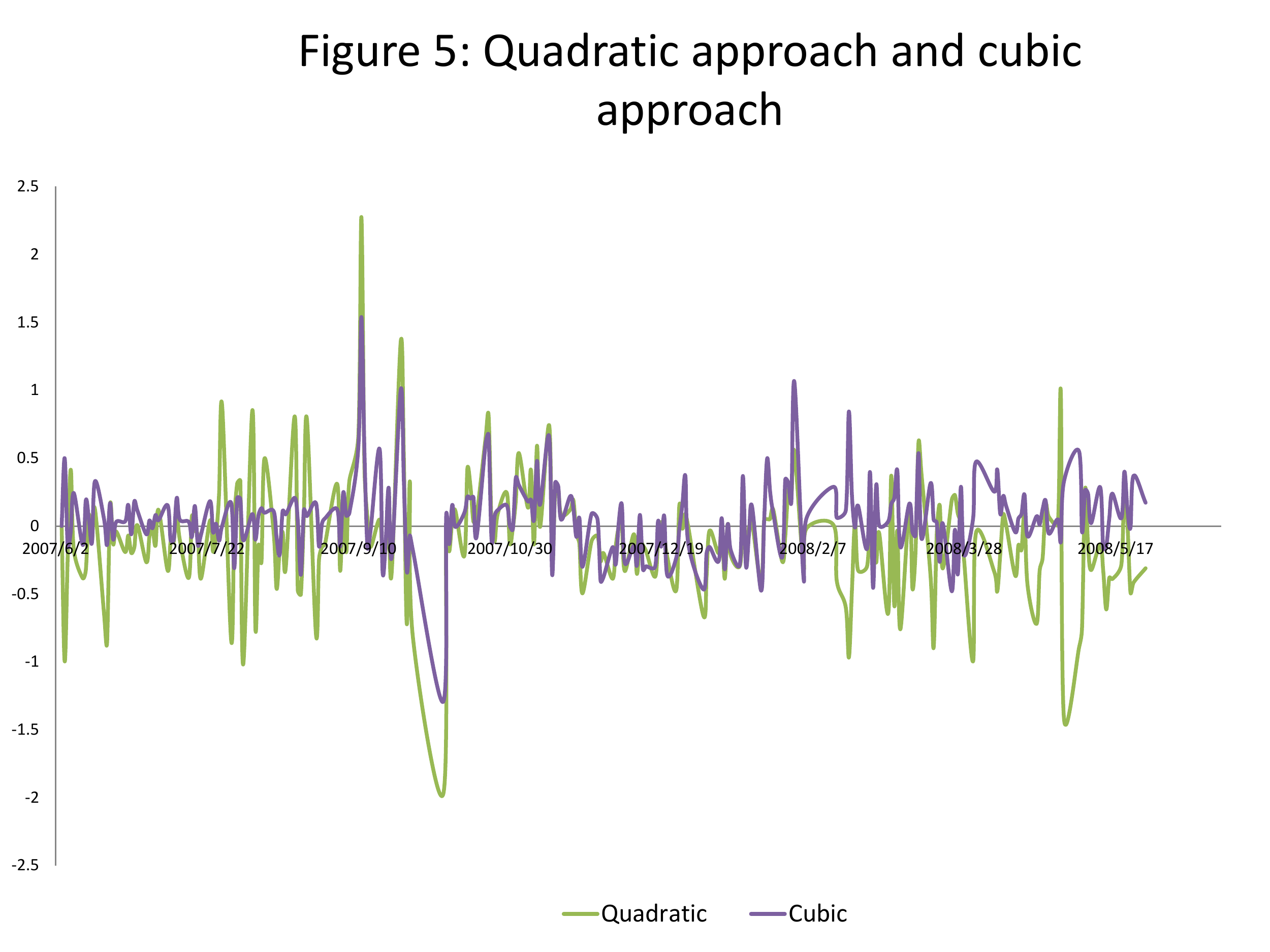}
\label{hedge3}
\end{center}
\end{figure}

\section{Conclusion}
To hedge against the interest-rate risk, one should describe the movement of the interest-rate term structure. The simplest approach is called the duration approach, which approximate the movement as a translation. This method needs only one instrument to hedge against the interest rate, and it is still widely used in the financial field.

We propose a new method to describe the movement of the yield-curve. Since the interest-rate term structure is smooth in maturity $T$ and irregular in time $t$, we can quantify the movement of the term structure as a function of $T$ and $t$. We use the polynomial interpolation method describe the yield-curve between $T_A$ and $T_B$, then the irregular movement with $t$ is the risk that should be hedged against. If we have two suitable hedging instruments on hand, we can use the quadratic-polynomial interpolation, which will describe the translation and the rotation of the term structure. If we have three suitable hedging instruments on hand, we can use the cubic-polynomial interpolation, which will describe the translation, the rotation and the twist of the term structure. For more complicated movement of the term structure, we can combine the traditional duration-convexity approach and the polynomial interpolation approach, but the shortage is that we have to use more than three hedging instruments, which would cause more risks such as the liquidity risk and the basis risk and lead less efficiency.

The empirical analysis shows that our hedging strategies are comparable or better than the traditional duration-convexity strategy. But all these methods will lose efficacy when hedging against the ultra-short-term interest-rate risk. Furthermore, We note that none of these approaches has the capability to deal with a sudden jump in the term structure, so we needs further researches.


\end{document}